\documentclass[a4paper,11pt]{article}
\usepackage{pos}
\usepackage{cleveref}
\usepackage{pgf}
\usepackage{booktabs}
\usepackage{array}
\usepackage{algorithm}
\usepackage{algpseudocode}

\newcolumntype{C}[1]{>{\centering\arraybackslash}m{#1}}
\newcolumntype{R}[1]{>{\raggedleft\arraybackslash}m{#1}}

\algnewcommand{\Input}{\item[\textbf{Input:}]}
\algnewcommand{\Parameters}{\item[\textbf{Parameters:}]}
\algnewcommand{\Output}{\item[\textbf{Output:}]}
\algrenewcommand\alglinenumber[1]{\scriptsize #1}

\renewcommand{\printHeadAuthors}{M. Maležič and J. Ostmeyer}

\title{Reducing the Gate Count with Efficient\\ Trotter-Suzuki Schemes}
\ShortTitle{Reducing the Gate Count with Efficient Trotter-Suzuki Schemes}

\author*[a]{Marko Maležič}
\author[a]{Johann Ostmeyer}

\affiliation[a]{Helmholtz institute for Radiation and Nuclear Physics, University of Bonn,\\
Nussallee 14-16, Bonn, Germany}

\emailAdd{malezic@hiskp.uni-bonn.de}
\emailAdd{ostmeyer@hiskp.uni-bonn.de}

\abstract{
  Hamiltonian formulations of lattice field theories provide access to real-time dynamics, but their simulation is difficult to implement efficiently.
  Trotter-Suzuki decompositions are at the center of time evolution computation, either on quantum hardware or classically, for instance with the use of tensor networks.
  While low-order Trotterizations remain the standard choice due to their simplicity, higher-order schemes offer the potential for improved efficiency.
  In this work we outline a short guide to Trotter-Suzuki schemes and their implementations in general.
  To help with this, we highlight new efficient schemes found by our optimization framework, and demonstrate their performance on the Heisenberg model.}

\FullConference{The 42nd International Symposium on Lattice Field Theory (LATTICE2025)\\
2-8 November 2025\\
Tata Institute of Fundamental Research, Mumbai, India\\}

\begin{document}
\maketitle


\section{Introduction}\label{sec:intro}

Different formulations of lattice field theory provide complementary insights into the properties of quantum field theories.
Due to the limitations of the Lagrangian formalism on the lattice, one may instead adopt the Hamiltonian formalism, which has recently seen significant developments~\cite{Jakobs:2025rvz, Kane:2025ybw, Fontana:2024rux, Jakobs:2025zcv}.
A key advantage of this approach is that it provides access to real-time evolution.
The dynamics are governed by the time evolution operator $U (-it) = e^{-i t H}$, defined as an exponentiation of the Hamiltonian $H$ for some time $t$.
Since Hilbert spaces in lattice theories typically grow exponentially, exact diagonalization of such Hamiltonians becomes infeasible.
Fortunately, the parts that compose a lattice Hamiltonian usually consist of local operators, such as a plaquette terms.
This allows the splitting of the time evolution operator $U$ using Trotter-Suzuki schemes $S_{n} (h) = U(h) + \mathcal{O} (h^{n+1})$, which incur an error of order $n$ in the time-step size $h$.

Currently, low order $n \leq 2$ Trotterizations are preferred in simulations of Hamiltonian lattice theories, due to their simple implementation~\cite{Funcke:2023RU, hariprakash2024strategiessimulatingtimeevolution}, e.g.\ classically using tensor networks or on quantum hardware.
However, a few methods have been proposed to construct higher orders~\cite{Suzuki:1976be, YOSHIDA1990262, OMELYAN2003272}, and recently progress has been made towards understanding the error accumulation of Trotter schemes~\cite{PhysRevX.11.011020, schubert2023trotter, chen2024trottererrortimescaling, chen2024errorinterferencequantumsimulation}.
Combining these, we built a general framework for constructing higher order schemes in our recent work~\cite{maležič2026efficienttrotter}, where we find novel schemes, which are believed to improve the efficiency of time evolution simulations.
With this work, we hope to provide a short guide to implementing time evolution of a desired system using a general Trotter-Suzuki scheme at any order $n$.
To demonstrate this, we recommend schemes at order $n = 4, 6$, which are believed to perform well in general, and we show their improved efficiency compared to historical schemes on the Heisenberg XXZ model.


\section{A guide to Trotter-Suzuki decomposition schemes}\label{sec:guide}

We begin by introducing the necessary concepts and definitions for discretized time evolution.
Our goal is not to improve the accuracy of a single step, but rather to approximate the full time-evolution operator $U(t)$ at a fixed total time $t$.
To this end, we present a general Trotter-Suzuki formula, briefly explain the methods to obtain scheme parameters, and write down the algorithm, which implements a general evolution in time.

\subsection{Theoretical foundations}

\begin{figure}[h!]
	\begin{center}
		\includegraphics[width=0.99\textwidth]{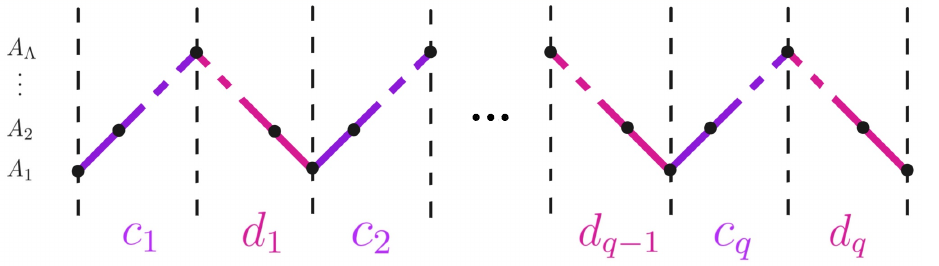}
	\end{center}
	\caption{Representation of a Trotter-Suzuki scheme with an arbitrary number of stages $\Lambda$.
		There are $q$ cycles, each consisting of two ramps.
		Ramps forward are indicated by the purple line, while pink lines represent the backward ramps.
		Read from either side, while multiplying exponents of operators $A_{k}$ with appropriate parameters $c_{i}$ or $d_{i}$, one obtains the decomposition from Eq.~\eqref{eq:general_scheme}.
		Adapted from~\cite{Ostmeyer:2022}.}\label{fig:ramp}
\end{figure}

Trotter-Suzuki schemes are approximations of operator exponentials~\cite{lie1888theorie, trotter_original}, such as the time evolution operator $U(-i t) = e^{-i t H} \approx \left[ S_{n} (h) \right]^{t/h}$, with a decomposition $S_{n} (h)$ of order $n$, which splits it into $N_{t} = t/h$ time steps of length $h$.
Each step generates a local error of order $\mathcal{O} (h^{n+1})$, which accumulates through time and results in the following global error $\mathcal{O} (h^{n+1} N_{t}) = \mathcal{O} (t h^{n})$.
Now we consider a general Hamiltonian $H$ as a sum of $\Lambda$ local non-commuting terms $A_{k}$: $H = \sum_{k}^{\Lambda} A_{k}$.
A scheme is composed of many \textit{stages}: $e^{c_{i} h A_{k}}$ and $e^{d_{i} h A_{k}}$, over \textit{sub-steps} $c_{i} h$ and $d_{i} h$, where the scheme parameters $c_{i}$ and $d_{i}$ define the specific decomposition $S_{n} (h)$ in the step size $h$.
The stages create either a ramp forward $\prod_{k=1}^{\Lambda} e^{c_{i} h A_{k}}$ or a ramp backward $\prod_{k=\Lambda}^{1} e^{d_{i} h A_{k}}$ (take care with the order of the indices $k$).
Together a ramp forward and backward complete a \textit{cycle} and a single step is composed of $q$ cycles.
This quantity groups the schemes of same $q$, and defines their order $n$.
In \Cref{fig:ramp} we visualize these concepts, which produce a general formula for a decomposition scheme:

\begin{equation}
	S_{n} (h) = \left( \prod_{k = 1}^{\Lambda} e^{c_{1} h A_{k}} \right) \left( \prod_{k = \Lambda}^{1} e^{d_{1} h A_{k}} \right) \cdots \left( \prod_{k = 1}^{\Lambda} e^{c_{q} h A_{k}} \right) \left( \prod_{k = \Lambda}^{1} e^{d_{q} h A_{k}} \right), \label{eq:general_scheme}
\end{equation}

Within an order $n$ and cycle $q$, there can exist many valid solutions for a decomposition scheme, but they vary according to their efficiency $\mathrm{Eff}_{n}$.
A scheme is efficient, if its leading order errors $\textrm{Err}_{n}$ are small compared to the number of cycles $q$ it requires, which can be defined in the following way,

\begin{equation}
  \textrm{Eff}_{n} = \frac{1}{q^{n} \textrm{Err}_{n}}.
\end{equation}

\noindent
For further details on how leading order errors $\textrm{Err}_{n}$ can be defined and computed refer to~\cite{OMELYAN2003272, maležič2026efficienttrotter}.
While efficiency is not comparable between orders, it is comparable between cycles $q$ within an order $n$.
Construction of high order $n \geq 4$ schemes historically started by taking schemes of lower order, e.g. the Verlet or Leapfrog scheme~\cite{Verlet:1967} at $n = 2$, as building blocks to obtain a higher order.
Methods that relied on these were derived by Suzuki~\cite{Suzuki:1976be, Hatano_2005} and Yoshida~\cite{YOSHIDA1990262}.
They however, fail to produce efficient schemes, which decreases their usability.

\begin{figure}[h]
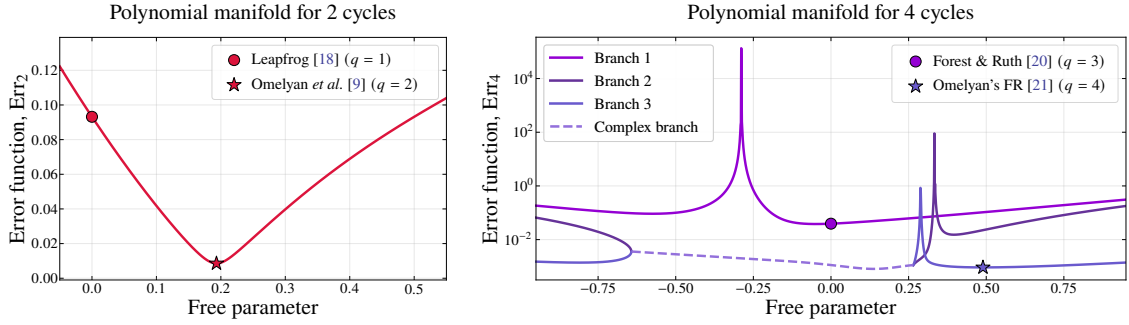

	\begin{center}
    \resizebox{0.405\textwidth}{!}{%
		\input{Figures/Manifold_q2.pgf}
	  }
    \resizebox{0.585\textwidth}{!}{%
		\input{Figures/Manifold_q4.pgf}
	  }
	\end{center}
	\caption{Error manifolds of $2^{\textrm{nd}}$ order schemes at $q = 2$ cycles (left) and $4^{\textrm{th}}$ order schemes at $q = 4$ cycles (right).
		The error function for 2 cycles is a simple one, with a single minimum, which is not hard to minimize.
    We plot it as a star, as well as the Leapfrog scheme, which can be found at null free parameter.
		The picture is more complicated at $q = 4$ cycles, where one finds 3 branches, two of them merging into a complex-valued parameter region, and 6 minima in total.
    We plot the global minimum again, and present the scheme by Forest \& Ruth on the real branch, where the value of the free parameter reaches zero.
    More manifold visualizations at $q = 5$ and $q = 6$ cycles are available in our repository~\cite{markomalezic_2026_18347430}.}\label{fig:manifolds}
\end{figure}

A less trivial method to construct schemes from scratch was pioneered by Omelyan \textit{et al.}~\cite{OMELYAN2003272}, which we extended into a general framework in our work~\cite{maležič2026efficienttrotter}.
There we focused on symmetric schemes, which produce even orders $n = 2, 4, 6, \ldots$, and are preferred due to their improved efficiency.
We were able to identify the polynomial structure of the leading order errors $\textrm{Err}_{n}$ in the scheme parameters for orders $n \leq 6$, which allowed us to directly impose the order constraints, and minimize the leading order error manifold.
Without delving into the details we visualize two of these polynomial manifolds in \Cref{fig:manifolds}.
The complexity of the manifold increases at higher cycles $q$ and order $n$, with more branches and minima emerging, which complicates the minimization.

\newpage

To give a rough understanding of how the number of constraints rises with the order, we present \Cref{tab:orders}.
We also show the number of scheme parameters, which scales with the number of cycles as $q + 1$, because of our restriction on symmetric schemes.
From these values we can calculate the number of free parameters at each cycle $q$, where orders $n = 2, 4$ give valid solutions everywhere.
However, orders $n \geq 6$ also have regions without free parameters, but `accidental' solutions of desired order are still found.
This is believed to result from the non-orthonormal nature of the constraints.
Higher orders $n \geq 10$ have not been explored much, but are conjectured to have no free parameter region~\cite{maležič2026efficienttrotter}.

\begin{table}[h!]
	\centering
	\caption{The table collects how the number of constraints changes with the order $n$ and how this affects the number of free parameters a scheme has at a given number of cycles $q$ (the number of all scheme parameters goes as $q+1$).
		Order $n = 2, 4$ both have fewer constraints, which allows us to optimize the free parameters.
		Orders $n \geq 6$ have a region of cycles $q$, where the no.\ of constraints is larger than the no.\ of scheme parameters, but valid solutions of desired order can still be found.
    Parentheses denote ranges of integer values and information on which cycle $q$ corresponds to which order $n$ can be found in Table 1 of Ref.~\cite{BLANES2002313}.}
	\begin{tabular}{lcccccc}
		\toprule
		\multicolumn{1}{l}{Order $n$}        & \multicolumn{1}{c|}{2}      & \multicolumn{1}{c|}{4}      &                                  \multicolumn{2}{c|}{6}                                     & \multicolumn{2}{c|}{8}                        \\
		\multicolumn{1}{l}{No.\ constraints} & \multicolumn{1}{c|}{2}      & \multicolumn{1}{c|}{4}      &                                  \multicolumn{2}{c|}{10}                                    & \multicolumn{2}{c|}{28}                       \\
		\midrule
		\multicolumn{1}{l}{Cycles $q$}       & \multicolumn{1}{c|}{[1, 2]} & \multicolumn{1}{c|}{[3, 6]} & \multicolumn{1}{c|}{\enspace[7, 8]\enspace} & \multicolumn{1}{c|}{[9, 14]}  & \multicolumn{1}{c|}{[15, 26]} & \multicolumn{1}{c|}{[27, 30]} \\
		\multicolumn{1}{l}{No.\ parameters}  & \multicolumn{1}{c|}{[2, 3]} & \multicolumn{1}{c|}{[4, 7]} & \multicolumn{1}{c|}{\enspace[8, 9]\enspace} & \multicolumn{1}{c|}{[10, 15]} & \multicolumn{1}{c|}{[16, 27]} & \multicolumn{1}{c|}{[28, 31]} \\
		\midrule
    \multicolumn{1}{l}{Free parameters}  & \multicolumn{1}{c|}{[0, 1]} & \multicolumn{1}{c|}{[0, 3]} & \multicolumn{1}{c|}{0}      & \multicolumn{1}{c|}{[0, 5]}   & \multicolumn{1}{c|}{0}        & \multicolumn{1}{c|}{[0, 3]}                   \\
		\bottomrule
	\end{tabular}\label{tab:orders}
\end{table}

\subsection{Algorithmic implementation}

We now present a simple way to implement time evolution of a quantum state, by using a general Trotterization.
Once the scheme parameters $c_{i}$, $d_{i}$ are known, and the action of local operators $A_{k}$ on the state has been implemented, the ramp approach (see Eq.~\eqref{eq:general_scheme}) can be applied.
The pseudocode for such time evolution is given in \Cref{alg:trotter}, where it is clear that the computational cost scales with the number of steps $N_{t}$ and the number of cycles $q$.
This algorithm does not include the grouping of stages with the same operator $A_{1}$ or $A_{\Lambda}$, but different parameter $c_{i}$ and $d_{i}$, which happens where two ramps meet.
Implementing this within a single step reduces the number of operations by $2q - 1$, while expanding it to the full evolution gives another reduction of $N_{t} - 1$.
While scheme parameters are readily available in literature or at our repository~\cite{markomalezic_2026_18347430}, the tricky part remains the implementation of operators $A_{k}$ in our models.
These depend on the system and the nature of our simulations, whether this includes tensor network manipulation or gate implementation on quantum circuits.
An elegant study about how this could be implemented on quantum hardware for lattice gauge theories is presented in Ref.~\cite{Davoudi:2022xmb}

\begin{algorithm}
\caption{Implementation of time evolution for a general Trotter-Suzuki scheme, based on the ramp approach~\eqref{eq:general_scheme}.}\label{alg:trotter}

\begin{algorithmic}[1]   
\Input Initial state $|\psi_{0}\rangle$, scheme parameters $c_{i}$, $d_{i}$ and operators $A_{k}$
\Parameters Total time $t$, number of time steps $N_{t}$
\Output Final state $|\psi\rangle$

\State $|\psi\rangle \gets |\psi_{0}\rangle$
\State $h \gets t / N_{t}$

\For{$\textrm{step} = 1, 2, \dots, N_{t}$}
  \For{$i = q, q-1, \ldots, 1$}
    \For{$k = 1, 2, \ldots, \Lambda$}
      \State $|\psi\rangle \gets e^{d_{i} h A_{k}} |\psi\rangle$
    \EndFor
    \For{$k = \Lambda, \Lambda - 1, \ldots, 1$}
      \State $|\psi\rangle \gets e^{c_{i} h A_{k}} |\psi\rangle$
    \EndFor
  \EndFor
\EndFor

\end{algorithmic}
\end{algorithm}


\section{Heisenberg model -- a practical example}\label{sec:practical}

We now apply the principles of the guide in \Cref{sec:guide} to provide novel schemes at orders $n = 4, 6$ with improved efficiency, and show their practical performance on the Heisenberg model.
Through our framework we were able to identify the lowest lying minima of the leading order error manifolds, i.e.\ the top most efficient schemes at each number of cycles $q$.
In practice however, this alone does not imply minimal Trotter error, because of other effects which influence error accumulation with time, some of which are discussed in~\cite{schubert2023trotter, chen2024trottererrortimescaling}.
An important result which we also observed, is the fact that the values of scheme parameters $c_{i}$ and $d_{i}$ can influence error accumulation.
Specifically, the further the values are from the \textit{origin point} $\bar{x} = \frac{1}{2 q}$, the worse the performance of a scheme.
Taking this into account as well as the theoretical efficiency, we recommend the use of scheme parameters found in Tables~\ref{tab:recommened6} and \ref{tab:recommened14} for orders $n = 4, 6$ respectively.
We find that maximal efficiency is found at maximal number of cycles, which results in a $q = 6$ scheme for order $n = 4$, and a $q = 14$ scheme for order $n = 6$.
Noteworthy is that both of these schemes are $1^{\textrm{st}}$ local minima of the error manifolds, not the global ones.
However, they are also as close as a scheme can be to the origin point $\bar{x}$, which improves the error accumulation.
Although, these two schemes are believed to perform well consistently across different models, it is possible that some other schemes work particularly well in a system of interest.
For this reason, we provide all found schemes at orders $n \leq 6$ in our repository found at~\cite{markomalezic_2026_18347430}, and now give a short example on how to work with them.

\begin{table}[h!]
	\centering
	\begin{minipage}[t]{0.48\textwidth}
		\centering
		\captionof{table}{Recommended parameters $c_{i}=d_{q+1-i}$ for an order $n = 4$ scheme at $q = 6$ cycles.
                      The scheme has the second-highest efficiency at this order, and performs well due to its proximity to the origin point.}\label{tab:recommened6}
		\vspace{2.0cm}
		\begin{tabular}{C{0.5cm}C{5.75cm}}
			\toprule
			$i$ &                   $c_{i} = d_{q+1-i}$   \\
			\midrule
			1  &                  $0.074082572180463262$  \\
			2  &                  $0.232923088374338803$  \\
			3  &                  $0.296820560634668408$  \\
			4  &                  $0.122086989386933251$  \\
			5  & \hspace{-0.45cm} $-0.350153632343424469$ \\
			6  &                  $0.124240421767020743$  \\
			\bottomrule
		\end{tabular}
	\end{minipage}
	\hfill
	\begin{minipage}[t]{0.48\textwidth}
		\centering
		\captionof{table}{Recommended parameters $c_{i}=d_{q+1-i}$ for an order $n = 6$ scheme at $q = 14$ cycles.
                      The scheme has the second-highest efficiency at this order, and performs well due to its proximity to the origin point.}\label{tab:recommened14}
		\begin{tabular}{C{0.5cm}C{5.75cm}}
			\toprule
			$i$ &                   $c_{i} = d_{q+1-i}$   \\
			\midrule
			1  &                  $0.037251326545569924$  \\
			2  &                  $0.120600278793781562$  \\
			3  &                  $0.266062994460763541$  \\
			4  &                  $0.163668553338143183$  \\
			5  &                  $0.071316838327437583$  \\
			6  &                  $0.058117508592333414$  \\
			7  &                  $0.188707697234255120$  \\
			8  & \hspace{-0.45cm} $-0.200016005078878524$ \\
			9  &                  $0.074145714537530386$  \\
			10 &                  $0.087345801243357893$  \\
			11 &                  $0.044234977360777830$  \\
			12 & \hspace{-0.45cm} $-0.230821838291030424$  \\
			13 & \hspace{-0.45cm} $-0.237197828922049295$  \\
			14 &                  $0.056583981858007803$  \\
			\bottomrule
		\end{tabular}
	\end{minipage}
\end{table}

For our example, we choose the Heisenberg XXZ model due to its straightforward application on quantum computers.
The Hamiltonian is defined over a periodic spin chain of length $L$:

\begin{equation}
	H = \sum_{i = 1}^{L} \left(\sigma_{i}^{x} \sigma_{i+1}^{x} + \sigma_{i}^{y} \sigma_{i+1}^{y} + \sigma_{i}^{z} \sigma_{i+1}^{z} + h_{i} \sigma_{i}^{z} \right),
\end{equation}

\noindent
where $\sigma_{i}^{\alpha}$ denote Pauli matrices, which are naturally implementable on quantum circuits as local gates.
The magnetic field $h_{i} \in [-0.1, 0.1]$ is sampled randomly from a uniform distribution.
The model can be diagonalized exactly for short chains $L \lesssim 20$, which allows the comparison of the time evolution operator $U$ with the Trotterized approximation $S_{n} (h)^{t/h}$.
We estimate the Trotter error by using the Frobenius norm,

\begin{equation}
	\Delta_{n}^{\textrm{exp}} = \frac{1}{\sqrt{N}} \left\| U (t) - S_{n}(h)^{t/h} \right\|_{F} = \frac{1}{\sqrt{N}} \sqrt{\sum_{v} \left| U(t) \cdot v - S_{n}(h)^{t/h} \cdot v \right|^{2}}, \label{eq:Frobenius}
\end{equation}

\noindent
where the sum runs over the basis states $v$ of the corresponding $N$-dimensional vector space with $N = 2^{L}$ and the $| \cdot |$ represents the Euclidean norm.
Besides the Frobenius norm other approximations like the spectral norm or the error of the eigenvalues could also be used~\cite{morales2022greatly}.
We also need to decide on the operator ordering of our splitting, and for this example we choose the following,

\begin{equation}
  \begin{aligned}
    S^{(3)} (ih) &= \left( \prod_{i=1}^{L} e^{ih H_{i}^{x} c_{1}} \right) \left( \prod_{i=1}^{L} e^{ih H_{i}^{y} c_{1}} \right) \left( \prod_{i=1}^{L} e^{ih H_{i}^{z} c_{1}} \right) \cdots \\
    &\times \left( \prod_{i=1}^{L} e^{ih H_{i}^{x} d_{q}} \right) \left( \prod_{i=1}^{L} e^{ih H_{i}^{y} d_{q}} \right) \left( \prod_{i=1}^{L} e^{ih H_{i}^{z} d_{q}} \right).
  \end{aligned}
\end{equation}

\noindent
where we defined the local operator $H_{i}^{\alpha} = J^{\alpha} \sigma_{i}^{\alpha} \sigma_{i+1}^{\alpha} + \delta_{a z} h_{i} \sigma_{i}^{z}$.
By counting the number of operators needed for a single ramp, we see that we need $3$ such stages due to commutation relations between the operators.
Ordering may impact the efficiency of the simulation, and the optimal choice might vary with model and geometry of the system~\cite{schubert2023trotter}.

\begin{figure}[h!]
	\begin{center}
    \resizebox{0.99\textwidth}{!}{%
		\input{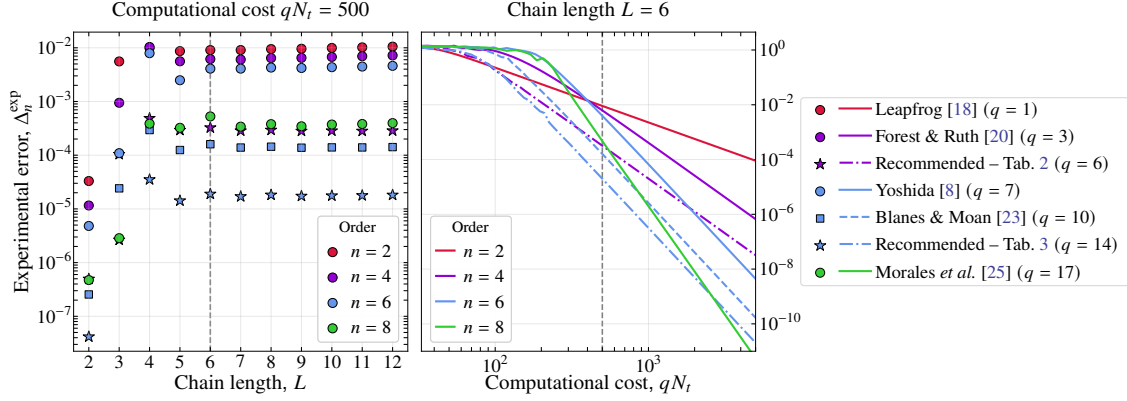}
	  }
	\end{center}
	\caption{Presented are the errors of Trotterized time evolution in the Heisenberg XXZ model approximated by the Frobenius norm $\Delta_{n}^{\textrm{exp}}$ (see Eq.~\eqref{eq:Frobenius}), for a collection of historical Trotterizations and our two recommended schemes at orders $n = 4, 6$ (see Tab.~\ref{tab:recommened6} and \ref{tab:recommened14}).
           On the left-hand side we observe the error as a function of the system size $L$, and find that it plateaus towards the thermodynamic limit.
           We present the improved efficiency of our novel schemes with respect to the computational cost $q N_{t}$, and find that our $n = 6$ scheme performs better than the historical schemes in a large region of the cost.
           The data was simulated at total time $t = 10$, and the gray lines indicate where the two plots were simulated at, i.e. $q N_{t} = 500$ and $L = 6$.}\label{fig:Trotter_error}
\end{figure}

With the model defined and prepared for simulation, we now show the performance and some properties of our novel schemes, in order to guide their use in other systems of interest.
Firstly, we can learn a lesson by simulating different chain lengths $L$ at a fixed computational cost $q N_{t} = 500$ and total time $t = 10$.
We plot the Trotter error $\Delta_{n}^{\textrm{exp}}$ for a collection of historical schemes and our recommended $n = 4, 6$ schemes on the left of \Cref{fig:Trotter_error}.
As one might expect, short chains exhibit unpredictable fluctuations.
However, these plateau around $L \approx 5$ in the Heisenberg model, and we observe a constant value, which is expected to extend into the thermodynamic limit across all schemes.
This is a universal feature, since the schemes were constructed in a model-agnostic way.
This opens a path to model-dependent tuning, where a small system with a known exact solution can be used as a benchmark for Trotter schemes.
Once the optimal one is found, it can be used reliably at larger system size, where no reference point is available.

Secondly, on the right-hand side of \Cref{fig:Trotter_error} we present the improved performance of our novel schemes compared to historical ones across the computational cost $q N_{t}$.
For this simulation we fix the system size in the plateau at $L = 6$, and compare the accumulated error $\Delta_{n}^{\textrm{exp}}$ at fixed total time $t = 10$.
On the lower end of the cost we find a plateau, where we approach the theoretical limit of the Frobenius norm.
Afterwards, we observe the scaling region, which follows the scaling law $\mathcal{O} (h^{n}) = \mathcal{O} (N_{t}^{-n})$.
Importantly, our recommended $6^{\textrm{th}}$ order scheme at $q = 14$ cycles (see Tab.~\ref{tab:recommened14}) performs better than historical schemes, even at lower costs, where the Leapfrog was thought to excel.
Furthermore, there is a significant region in the cost, where the best order $n = 8$ schemes by Morales $\textit{et al.}$~\cite{morales2022greatly} perform worse.
Using our framework, we believe the efficiency could still be improved at order $n = 8$.

\newpage


\section{Conclusion}\label{sec:conclusion}

To sum up, we presented a concise manual on Trotter-Suzuki decompositions schemes.
In particular, we focused on improving time evolution algorithms at a fixed total time, rather than optimizing individual Trotter steps.
In \Cref{sec:guide}, we offered a brief theoretical guide to general Trotterizations and some of their properties.
We also provided some simple code to implement Trotterized time evolution at any order $n$ and cycle $q$ (see \Cref{alg:trotter}).
This is followed by \Cref{sec:practical}, where we discussed novel efficient schemes and presented their performance on the Heisenberg model (see \Cref{fig:Trotter_error}).
Using this practical example, we demonstrated that schemes analyzed at small system sizes can be reliably applied to larger systems.

\section*{Code and Data}

All the code needed to reproduce the results in this work can be found on Github or Zenodo~\cite{markomalezic_2026_18347430}.
Time evolution routines for the dynamics of the Heisenberg model are written is \texttt{C++}, with a dependency on Eigen~\cite{eigenweb}.
The data for the time evolution can be reproduced with the given code, but it is slightly too large for publication on our repository.
Nonetheless, the data will be gladly provided upon request.

\newpage

\section*{Acknowledgements}
We thank Anthony Kennedy, Paul Ludwig, Emanuele Mendicelli, Benjamin Søgaard, Lorenzo Verzichelli and Jesse Stryker for insightful discussions.
The authors gratefully acknowledge the access to the Marvin cluster of the University of Bonn.
This work was funded by the Deutsche Forschungsgemeinschaft (DFG, German Research Foundation) as part of the CRC 1639 NuMeriQS – Project number 511713970.

\bibliographystyle{JHEP}
\bibliography{bibliography}

\end{document}